%% file: main.tex
\documentclass[conference]{IEEEtran}
\usepackage{xcolor}
\definecolor{dkgreen}{rgb}{0,0.6,0}
\usepackage[dvipsnames,table]{xcolor}
\usepackage{tikz}

\usepackage{stfloats}
\usepackage{amssymb}
\usepackage{graphicx}
\usepackage{amsmath}
\usepackage{amssymb}
\usepackage{listings}
\usepackage{caption}
\usepackage{subcaption}
\usepackage{tikz}
\usetikzlibrary{shapes,arrows,positioning}
\usepackage{hyperref}
\usepackage{orcidlink}
\usepackage{stfloats}
\usepackage{enumitem}
\usepackage{stfloats}
\usepackage{multirow}
\usepackage{hhline}
\usepackage{array}
\usepackage{tabularx}
\usepackage{subcaption}
\usepackage{amsmath,amsfonts}
\usepackage{bbold}
\usepackage{listings}
\usepackage{multirow}    % for multirow if needed
\usepackage{colortbl}
\usepackage{booktabs}
\usepackage{caption}
\usepackage{graphicx}
\usepackage{array}
\usepackage{makecell}
\usepackage{fixme}
\renewcommand{\thetable}{\arabic{table}}
\ifCLASSOPTIONcompsoc
  \usepackage[nocompress]{cite}
\else
  \usepackage{cite}
\fi

\ifCLASSINFOpdf
\else
  
\fi 
\usepackage{braket}
\usepackage{float}
\usepackage{graphicx}
\usepackage{xcolor}

\usepackage{amssymb}
\usepackage{pifont}
\usepackage{xspace}
\newcommand{\ignore}[1]{}

\newcommand{\Code}[1]{\lstinline{#1}}

\usepackage{makecell}

\definecolor{aliceblue}{rgb}{0.94, 0.97, 1.0}
\usepackage{xcolor}
\usepackage{tcolorbox}
\tcbset{width=1\columnwidth,boxrule=1pt,colback=aliceblue,arc=1pt,left=2pt,right=2pt,boxsep=0.5pt}

\def\BibTeX{{\rm B\kern-.05em{\sc i\kern-.025em b}\kern-.08em
    T\kern-.1667em\lower.7ex\hbox{E}\kern-.125emX}}

\usepackage[flushleft]{threeparttable}
\newcommand{\sys}{{\tt QEX}\xspace}
\newcommand{\sysh}{{\tt QEX-H}\xspace}

\newcommand{\eg}{\textit{e.g.}}
\newcommand{\ie}{\textit{i.e.}}

\author{
\IEEEauthorblockN{Yicheng Guang, Pietro Zanotta, Kai Zhou, Yueqi Chen, Ramin Ayanzadeh}\vspace{-0.5em} \\
\IEEEauthorblockA{\textit{University of Colorado Boulder}, Boulder, USA}\vspace{-0.8em} \\
\IEEEauthorblockA{\{yicheng.guang, pietro.zanotta, kai.zhou-1, yueqi.chen, ayanzadeh\}@colorado.edu}
}

\begin{document}

\title{
  On the Potential of Quantum Computing in 
  Classical Program Analysis
  }

\IEEEtitleabstractindextext{%

\input{./sections/0Abstract}

 
 }

\maketitle

\IEEEdisplaynontitleabstractindextext

\IEEEpeerreviewmaketitle

\input{./sections/intro}
\input{./sections/background}
\input{./sections/qex}

\input{./sections/qex-h}

\input{./sections/related}
\input{sections/conclude}
\newpage

\appendix
\input{sections/Appendix}

\newpage
\bibliographystyle{IEEEtran}
\bibliography{ref}

\end{document}

%% file: sections/0Abstract.tex
\begin{abstract}

Classical program analysis techniques, such as abstract interpretation and symbolic execution, are essential for ensuring software correctness, optimizing performance, and enabling compiler optimizations. 
However, these techniques face computational limitations when analyzing programs with large or exponential state spaces, limiting their effectiveness in ensuring system reliability. 
Quantum computing, with its parallelism and ability to process superposed states, offers a promising solution to these challenges.
In this work, we present \emph{\sys{}}, a design that uses quantum computing to analyze classical programs. 
By synthesizing quantum circuits that encode program states in superposition and trace data dependency between program variables through entanglement, \sys{} enables the simultaneous exploration of program behaviors, significantly improving scalability and precision. 
This advancement has broad applications, from debugging and security verification to optimizing compilers for next-generation hardware. 
As a proof-of-concept, we evaluated \sys on 22 benchmark programs, demonstrating its effectiveness in analyzing program states.
% #RA, check the following sentence for correctness
To support more language features and make \sys realized sooner in Fault-Tolerant Quantum Computing (FTQC), we propose \sysh which hybridizes \sys with classical analysis techniques.
%#RA it would be better, if we have space, in one short sentence (similar to "As a proof-of-concept ...",[above]), we should compare results with \sys 
To our knowledge, this work is the first proposal to use quantum computing for classical program analysis.

\end{abstract}

%% file: sections/intro.tex
\section{Introduction}

% {\textcolor{red}{TODO: Delete realiability and emphasize the importance of program analysis}}
% \todo{prior works proposing adder doesn't mean that's all for program analysis.}

Ensuring the reliability of systems—both software and hardware—is a critical goal in modern computing.
Program analysis is a foundational tool in achieving this by identifying bugs~\cite{bugfinding}, detecting vulnerabilities~\cite{vulnerability}, and optimizing overall system performance~\cite{optimization}.
It is also essential for compiler optimizations~\cite{haghighat1992symbolic}, such as moving loop-invariant computations outside loops~\cite{si2018learning}, improving program efficiency.
These applications collectively enhance software quality and ensure robust behavior in diverse environments.

However, program analysis is constrained by the undecidability~\cite{undecidability} of computing exact program semantics, requiring approximation techniques.
Over-approximation methods, like abstract interpretation~\cite{cousot1977abstract}, ensure scalability by considering a superset of possible behaviors, but often result in false positives.
Under-approximation methods like fuzzing~\cite{fuzztesting} focus on specific cases or paths, making them efficient but sacrificing soundness and potentially missing vulnerabilities.
Other approaches, like symbolic execution~\cite{king1976symbolic}, aim to balance accuracy and soundness but face challenges like path explosion and solver limitations.
% By simplifying program semantics into abstract domains, they guarantee soundness but often produce false positives. 
% On the other hand, under-approximation methods, such as \emph{Fuzz Testing} ~\cite{fuzztesting}, focus on specific cases or execution paths, which makes them computationally efficient but incomplete. This limitation leaves potential vulnerabilities undetected.

% Other approaches, like Symbolic Execution~\cite{king1976symbolic}, aim to balance these trade-offs by treating inputs as symbolic variables and exploring paths using logical constraints. 
% While theoretically promising, symbolic execution faces practical challenges like path explosion and the limited capability of constraint solvers which struggles with high computational complexity (\( O(N-M) \)). 
% N is the size of the whole searching space. M is the program states which satisfy the constraints. N will grow exponentially with the increase of the number of variables. T
% These issues restrict its applicability to only small programs and can lead to under-approximation in practice. Despite recent advancements like path pruning and concolic execution, symbolic execution can only work on small programs and remains limited when applied to real-world, large-scale systems.

Quantum computing introduces a transformative paradigm that leverages principles like quantum superposition and entanglement to provide computational advantages.
As quantum computing has demonstrated promising utility in multiple domains such as Finance~\cite{rainbow}, Biology~\cite{mrna}, and Chemistry~\cite{ground}, it is natural to ask whether these same benefits could extend to program analysis, thereby overcoming the limitations of classical techniques.

% especially in exploring exponentially large state spaces. For example, Grover’s algorithm ~\cite{grover1996fast} offers quadratic speedup for search problems, making it highly suitable for exhaustive program state exploration—a core challenge in program analysis. The ability to harness quantum computing for program analysis can significantly enhance the reliability of systems by addressing limitations of classical methods.

In this work, we propose \emph{\sys{}}, a novel quantum design that provides a new way to analyze classical programs, outperforming abstract interpretation and symbolic execution in several aspects.
\sys leverages quantum superposition to efficiently encode $2^{N}$ program states into $N$ qubits.
It then produces quantum circuits to interpret the semantics of various program statements, enabling the simultaneous exploration of the entire program state space.
Additionally, \sys leverages entanglement to track data dependencies between program variables, ensuring analysis accuracy.
The qubits are finally measured to decode program states of interest through a 
fixed-point quantum search~\cite{fixed-point} with an optimal number of queries.
Our experiment shows that \sys can effectively eliminate over- approximation and under-approximation compared to classical analysis techniques. 

% \todo{However, \sys can only be applied to WHILE language~\cite{while}. 
% Memory related operations can't be analyzed due to the lack of an ideal QRAM~\cite{qram0,qram1}. 
% On the other hand, although \sys is designed for fault-tolerant quantum computer~\cite{shor1996fault}, reducing resource consumption like qubits, gates and circuit depth to make our design realized in NISQ era~\cite{preskill2018quantum} is also important.}.

However, \sys cannot support pointer-related operations such as pointer assignments or dereferencing. 
Moreover, the additional physical qubits and gates required by Quantum Error Correction (QEC) in Fault-Tolerant Quantum Computing (FTQC) may postpone the realization of \sys.
To address this, we introduce \emph{\sysh}, a hybrid approach that combines \sys with classical methods. \sysh extends \sys to support more language features, thereby maximizing its overall utility. 
Furthermore, by substantially reducing circuit size and QEC overhead, \sysh makes \sys more attainable within the FTQC era while preserving the key quantum advantages.
% \todo{However, \sys isn't able to support memory-related operations like pointer assignments or dereferencing. Moreover, the additional physical qubits and gates required by Quantum Error Correction (QEC) in Fault-Tolerant Quantum Computing (FTQC) may postpone the realization of \sys. 
% To address this, we introduce \emph{\sysh}, a hybrid approach that combines \sys with classical methods. \sysh extends \sys to support more language features, thereby maximizing its overall utility. Furthermore, by substantially reducing circuit size and QEC overhead, \sysh makes \sys more attainable within the FTQC era while preserving the key quantum advantages.}

% While \sys is designed for Fault-tolerant quantum computation (FTQC), to make it practical in the Noisy Intermediate-Scale Quantum (NISQ) era so as to benefit program analysis sooner,
% % To drive the adoption of \sys in practice as quantum hardware continues to advance, 
% we further introduce a hybrid design \sysh which integrates \sys with classical techniques, enhancing its scalability and applicability.
% Through hybridization, \sysh reduces quantum resource consumption, supports more language features, and maximizes the utility of \sys.

\emph{To our knowledge, this is the first work to rigorously explore and validate the potential of quantum computing for classical program analysis}. Previous work like QCheck~\cite{guang2025quantum} only focuses on specific aspects with a lot of limitations.
Through our work, we aim to highlight the opportunities in this direction to invite future research as well as the challenges that must be overcome to achieve practical impact.
In summary, this work makes the following contributions:

% \vspace{0.05in}
% \noindent In summary, this work makes the following contributions:
\vspace{0.05in}
\begin{itemize}[itemsep=5pt]
    \item We introduce \sys, a novel quantum framework for program analysis. It leverages superposition to enable the simultaneous exploration of the entire program state space and utilizes entanglement to trace data dependencies among program variables. 
    % To our knowledge, this is the first work to leverage quantum computing for classical program analysis.     
    \item We demonstrate that \sys can effectively eliminate over/under-approximation, achieving greater accuracy and soundness compared to classical methods. 
    % \item We present \sysh, a hybrid design that integrate \sys with classical methods. It maximizes the utility of \sys while reducing the hardware requirements so that \sys can benefit program analysis sooner in the NISQ era.
    \item We present \sysh, a hybrid design that integrate \sys with classical methods. It maximizes the utility of \sys while reducing the hardware requirements in FTQC so that \sys can benefit program analysis sooner.
\end{itemize}

%% file: sections/background.tex
\section{Background and Motivation}
\label{sec:background}

In this section, we present the quantum background and quantum algorithms used in our design, followed by describing the limitations of classical program analysis techniques that motivate our design.

\subsection{Applications of Quantum Computing}
Quantum computing~\cite{quantumsurvey} is transforming various fields by solving complex problems beyond classical capabilities. 
In Finance, quantum algorithms improve derivative pricing, as seen in Quantum Amplitude Loading for Rainbow Options Pricing~\cite{rainbow}, which enhances multi-asset valuation using Iterative Quantum Amplitude Estimation. 
In Biology, quantum methods aid drug discovery, demonstrated in mRNA Secondary Structure Prediction~\cite{mrna}, where quantum optimization accurately predicts RNA structures. 
In Chemistry, quantum simulations accelerate material science, exemplified by Ground-State Energy Estimation~\cite{ground}, which optimizes density-functional theory for large-scale molecular modeling. 
These advancements highlight quantum computing’s growing impact across industries, promising breakthroughs in optimization, simulation, and data analysis. 
In Design Automation, Nils et al.~\cite{EquivalenceChecking} explore the possibility of a quantum solution to check the equivalence of classical circuits.

\emph{
Despite these wide applications in various domains, there is little work that comprehensively explore the potential advantages and challenges of applying quantum computing for classical program analysis.
}

\subsection{Quantum Algorithms and Circuits}
The foundation of quantum speedup lies in properties such as superposition—the ability of a qubit to exist in a combination of $|0\rangle$ and $|1\rangle$—and entanglement, where the state of two qubits are interdependent. 
Taking Grover's algorithm~\cite{grover1996fast} as an example, it is a widely studied quantum search algorithm that finds a target element in an unsorted database of \( N \) items, achieving quadratic speedup from the classical approach's   \( O(N) \) to  \( O(\sqrt{N}) \). 
Unlike classical algorithms that evaluate potential solutions one by one, Grover's algorithm~\cite{grover1996fast} is initialized with an uniform superposition of all possible states $|s\rangle$ to explore possible solutions simultaneously. 
The key component of the algorithm is amplitude amplification, which consists of an oracle operator $R_t=\mathbf I-2|t\rangle\langle t|$ and a diffusion operator $R_s=\mathbf I-2|s\rangle\langle s|$ with the target states $|t\rangle$ and the identity operator $\mathbf I$. 
$R_t$ flips the sign of the target states $|t\rangle$, effectively marking them from other states. 
$R_s$ then amplifies the probability of these marked states by flipping the sign of the initial state $|s\rangle$. 
Through iterative application of the Grover iterate $G=-R_sR_t$, the algorithm increases the probability of the target states
After approximately $O(\sqrt{N/M})$ iterations, the probability of observing a target state approaches 100\%, where $N$ is the number of possible solutions and $M$ is the number of target solutions.

Quantum algorithms like Grover's~\cite{grover1996fast} are implemented as quantum circuits composed of quantum gates. 
These gates perform unitary transformations to amplify the probability amplitude, with a measurement reading out the final states. 
The correct solution can be observed with a high probability. In practice, the circuits will be executed multiple times until the correct solution appears.

\subsection{The Fixed-point Quantum Search}

Grover's algorithm and its generalization, quantum amplitude amplification~\cite{amplitude} provides a quadratic speedup over classical algorithms.
However, prior knowledge of what fraction $M/N$ of the initial state is the target state is required, which is the so-called souffle problem. 
To overcome this limitation, some fixed point quantum algorithms have been proposed, in particular the $\pi/3$-algorithm~\cite{grover2005fixed}, which provides a lower bound on $M/N$, sacrificing the quadratic speedup. 

The work of~\cite{fixed-point}, on the other hand, provides the fixed-point behavior without sacrificing the quadratic speedup. For all $M/N\ge 1-\gamma^{2}$, this algorithm can extract the target state with the success probability $p_{L}\ge1-\delta^2$, where $L$ is the total number of Grover iterates, $\delta\in[0,1]$ is a parameter chosen by the user, and $\gamma=1/\cos[\arccos(1/\delta)/(2L+1)]$. 
Consequently, the complexity of the algorithm is $O\left(\log(2/\delta)\sqrt{N/M}\right)$, which reflects the quadratic speedup achieved.

\subsection{Limitations of Classical Program Analysis}
Despite the significance and utility of program analysis, its effectiveness is fundamentally constrained by the undecidability of computing a program’s exact semantics. Consequently, analysts must rely on either over-approximation or under-approximation techniques.
Over-approximation methods ensure scalability by considering a superset of all possible behaviors. 
They guarantee soundness but often lead to false positives, \ie, program states never occurring during actual execution. 
In contrast, under-approximation methods focus on specific cases or execution paths, making them computationally efficient but inherently incomplete.
Other approaches, like symbolic execution ~\cite{king1976symbolic}, attempt to balance these trade-offs.
However, challenges like path explosion and the computational complexity of constraint solving restrict their applicability to small programs and, in practice, can lead to under-approximation.
In the following, we discuss two representative analysis techniques, abstract interpretation and symbolic execution, to illustrate the inherent obstacles of classical analysis techniques and motivate the design of our quantum approach.

\vspace{5pt}
\noindent \textbf{Abstract Interpretation.}
This analysis framework maps concrete program states onto an abstract domain. 
Using transfer functions, the states evolve within this domain until reaching a fixed point.
For example, consider the assignment statement \Code{z := x + y;}. 
If \Code{x} can take values of either $1$ or $3$, an interval domain representing this range is $[1,3]$. 
Similarly, if \Code{y} can take values of either $2$ or $4$, the interval domain is $[2,4]$.
Applying the transfer function for addition, the interval domain of \Code{z} is computed as $[3,7]$. 
However, in an actual execution, \Code{z} can only take three values: $3$, $5$, and $7$, meaning that $4$ and $6$ are false positives introduced by the abstraction. 

These false positives are the cost of scalability, as abstract domains provide a more compact representation than enumerating individual states.
However, excessive false positives can reduce the usefulness of the analysis by overwhelming developers with false alarms.
Mitigating false positives typically involves refining abstractions or incorporating supplementary analyses, but these approaches increase computational complexity and may, in turn, compromise scalability.

\vspace{5pt}
\noindent \textbf{Symbolic Execution.}
This technique abstracts the input space of a program by representing program inputs as symbolic variables rather than concrete values.
As execution progresses, symbolic expressions track how these variables propagate through the program, forming a set of first-order constraints that define feasible execution paths. 
Constraint solvers, 
such as SAT/SMT solvers~\cite{de2008z3}, 
are then used to determine satisfiability, enabling the identification of feasible program behaviors and potential bugs.

However, this approach suffers from scalability challenges due to path explosion and the computational overhead of constraint solving, particularly in programs with loops, complex data structures, or high branching complexity.
For example, consider a program with $40$ \Code{if-else} statements, the total number of program paths is $2^{40}$.
Such state explosion growth quickly exceeds the memory capacity of classical computers.
Another major limitation of symbolic execution is the computational constraints of SAT/SMT solvers.
% \todo{RA said he want us to cite this (Symbolic execution for RSA), but I can't find paper talking about this. I can only cite paper talking about constraints for SMT solver}
Consider a program that implementing a RSA algorithm, solving the constraints generated by symbolically executing such a program is as difficult as breaking the RSA itself~\cite{SElimitations}.

%% file: sections/qex.tex
\section{\sys Design}
\label{sec:design}

As classical analysis techniques struggle to balance the exploration of a vast program state space with the accuracy and soundness needed to produce useful and reliable results, our work aims to explore the potential of quantum computing to address this challenge.

The superposition property of quantum computing offers a unique advantage in representing program states.
For a program with $n$ variables, each $m$ bits wide, its total state space comprises $2^{n\times m}$ possible combinations of values.
While this scale is intractable for classical computing, quantum systems can explore all program states simultaneously using $n \times m$ qubits in superposition, achieving an exponential reduction in resource requirements.

To exploit this advantage, we propose \sys which produces equivalent quantum circuits that explore the state space of a given program.
These circuits interpret the semantics of various program statements by applying amplitude amplification to the qubits that represent program variables and states.
The qubits are finally measured to decode feasible values a variable can take after program execution.
These decoded values, which reflect the program's behaviors, can be used to examine, for example, if a buffer index exceeds its boundary.

\begin{figure}[t]
    \centering
	\includegraphics[width=0.5\textwidth]{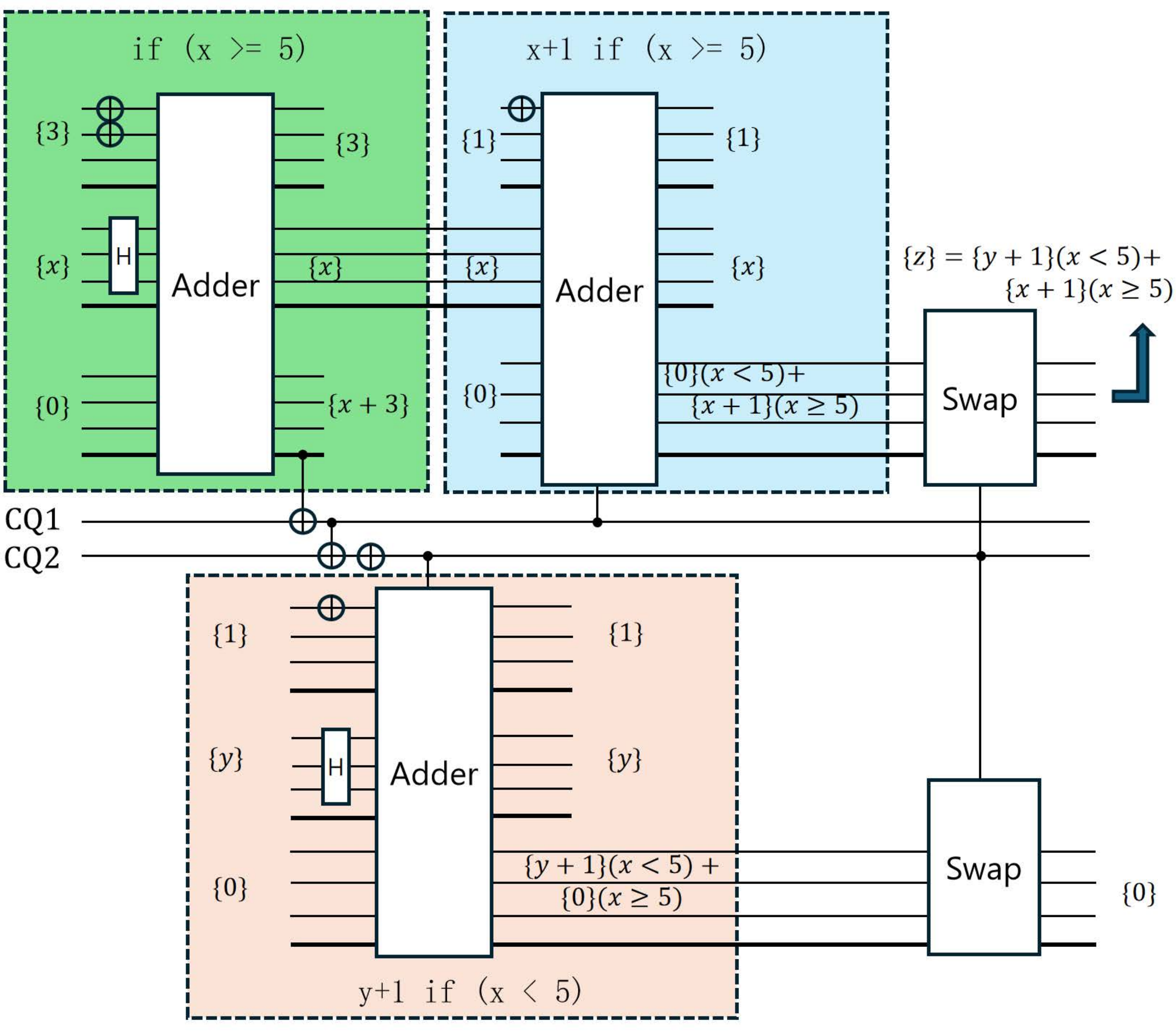}
    \caption{The unoptimized circuit to interpret an example program: \Code{if (x >= 5)\{z := x + 1;\} else \{z := y + 1;\}}. 
    $\{x\}$ and $\{y\}$ represents all possible input values of \Code{x} and \Code{y}.
    $\{y+1\}(x<5)+\{0\}(x\ge5)$ means corresponding qubits are measured as $\{y+1\}$ when \Code{x} is smaller than 5 and $\{0\}$ in the other situation.}
    \label{fig:ExampleCircuit} 
\end{figure}

\subsection{Design Overview}
\sys employs a universal gate set comprising U3 gates and control gates to construct circuits and analyze programs written in the WHILE language~\cite{while}. 
The WHILE language is an abstract programming language widely used in program analysis, formal verification, and theoretical computer science.
It is C-like, supporting assignment statements (\Code{x := a}) and conditional statements such as \Code{if} and \Code{while} with boolean predicates. 
Predicates include constants (\Code{true}, \Code{false}), negation, conjunction (\Code{and}), disjunction (\Code{or}), and relational operators (\Code{<, <=, >, >=}) applied to arithmetic expressions. Arithmetic expressions comprise variables, number literals, and operators (\Code{+,-,*,/}).

Figure~\ref{fig:ExampleCircuit} presents an unoptimized circuit for an example program: \Code{if (x >= 5)\{z := x + 1;\} else \{z := y + 1;\}}.
For illustration purpose, in this circuit, all variables are 3-bit wide with value ranging from $0$ to $7$, and are represented using three qubits plus a sign qubit (in bold).
Each input variable (\eg, \Code{x} and \Code{y}) is initialized in the state $|0\rangle$, followed by the application of a Hadamard gate to establish an equal superposition over all possible values.
Then, the circuit quantumly interprets the semantics of the program’s statements to amplify the amplitudes of different values the variable can take, thereby fully exploring its state space.  
For simplicity, in the figure, we use $\{x\}$  to indicate all possible values \Code{x} can take and $\{x+1\}(x<5)$ to represent that when \Code{x} is smaller than $5$, the values superpositioned in corresponding qubits are $\{x+1\}$.
Complete Dirac representations of initial state and final state are formulated in Equation~\ref{eq:xy_states} and ~\ref{eq:phi_state} in Appendix~\ref{appendix}.
Details are discussed below.

\subsection{Quantum Interpretation of Expressions and Statements}
\vspace{5pt}
\noindent \textbf{Arithmetic Expression.}
\label{sec:arithmetic}
For arithmetic expressions such as \Code{x + 1} and \Code{y + 1}, prior work~\cite{QuantumArithmeticTraditional} introduced a quantum adder design that replicates its classical counterpart. 
Following works~\cite{QuantumArithmeticFourier, QuantumArithmeticSignBit} further reduced the number of qubits and gates required by leveraging the Quantum Fourier Transform. 
However, these designs cannot be directly applied in \sys.
Technically, they require two sets of qubits as input operands, with one set simultaneously serving as storage for the arithmetic results.
Consequently, they permanently change the qubits' state, for example, from \Code{x} to \Code{x+1}.
When the value of \Code{x} is needed later, it has already been overwritten.
To address this issue, \sys extends the prior designs by incorporating additional CX gates to preserve the values of both input operands. 
As shown in Figure~\ref{fig:ExampleCircuit}, the extended adder outputs three sets of qubits: two sets encode the original values for the input operands, and the third set encodes the arithmetic result.
The extended designs for multipliers and dividers follow the same design.

% Here's an example of the function of adder ($U_1$):
% \begin{align*}
% U_1\left(\frac{\sqrt{2}}{2}(|00\rangle + |01\rangle) \otimes \frac{\sqrt{2}}{2}(|01\rangle + |10\rangle) \right) = \\
% \frac{1}{2}|0001\rangle + \frac{1}{2}|0010\rangle + \frac{1}{2}|0110\rangle + \frac{1}{2}|0111\rangle
% \end{align*}
% For this adder, it has two sets of qubits as input and one set of qubits will also be used as output of the addition results after the adder. If we want to keep both the original two sets of input qubits, we can extend the adder ($U_2$) like this:
% \begin{align*}
% U_2\left(\frac{\sqrt{2}}{2}(|00\rangle + |01\rangle) \otimes \frac{\sqrt{2}}{2}(|01\rangle + |10\rangle) \otimes |00\rangle\right) = \\
% \frac{1}{2}|000101\rangle + \frac{1}{2}|001010\rangle + \frac{1}{2}|010110\rangle + \frac{1}{2}|011011\rangle
% \end{align*}
% For this extended adder, we can keep both the inputs and assign the added output into a new set of qubits. The designs of multiplier and divider are like the extended adder. The results will be assigned to a new set of qubits.

\vspace{5pt}
\noindent \textbf{Assignment Statement.}
Logically, an assignment statement can be viewed as an addition where one operand is zero.
However, this interpretation is unnecessarily complex. 
Instead, we can simply use a CX gate to ``copy'' values between qubits that represent variables, thereby creating entanglement.
As will be discussed in Section~\ref{subsec:engtanglement}, this entanglement naturally preserves the desired data flow relationships for accurate program analysis.

% \todo{I define a new notation, add explanation in caption and add appendix for real entanglment states.}

\vspace{5pt}
\noindent \textbf{Conditional Statement.}
For conjunction (\Code{and}) and disjunction (\Code{or}) operations in predicates, they can be directly interpreted using corresponding quantum control gates such as Toffoli gate.
The challenge of conditional statements, however, lies in evaluating predicates to divide execution flow into different branches and subsequently merging the results after both branches are complete.

Figure~\ref{fig:ExampleCircuit} exemplifies how \sys structures signal qubits, control qubits, and swap gates to solve the problem.
First, the evaluation of the predicate \Code{x >= 5} is converted to an adder that adds $3$ to \Code{x}.
Assuming all variables in the example program are 3-bit width, this adding will overflow if \Code{x >=5}.
The resulting overflow sets the sign qubit (displayed in bold) to a determined $1$, which is then ``copied'' to the control qubit CQ1 through a CX gate.
CQ1 enables the circuit that performs \Code{x+1} in the true branch.
If no overflow occurs, the sign qubit is a determined $0$ which is ``copied'' first to CQ1 and then to CQ2.
Applying an X gate to CQ2 flips its state to a determined $1$ which enables the circuit that performs \Code{y+1} in the false branch.

Note that throughout the circuit's execution, the value of \Code{x} is in superposition, so are the values of control qubits CQ1 and CQ2.
Therefore, both the true and false branches are explored simultaneously, each corresponding to different conditions on \Code{x}.
The true branch produces the state (in the bottom of blue area of Figure~\ref{fig:ExampleCircuit}) is
\begin{equation}
   \{0\}(x<5)+\{x+1\}(x\ge5)
\end{equation}
indicating that measurement result is 0 when \Code{x} is smaller than 5 and $\{x+1\}$ otherwise.
Meanwhile, the false branch produces the state
\begin{equation}
    \{y+1\}(x<5) + \{0\}(x \geq 5)
\end{equation}
Finally, to merge the program states from the two branches, CQ2 enables a swap gate that exchanges $\{y+1\}$ and $\{0\}$ in the part of false branch \Code{x < 5}.
The measurement results of qubits representing z (qubits in the bottom of blue area) are 
\begin{equation}
    \{y+1\}(x<5)+\{x+1\}(x\ge5)
\end{equation}
which comprehensively represents all possible values that the variable \Code{z} can take after the program executes.

% All qubits in the circuit, except those storing the final \Code{z}, can be reused if there are subsequent statements in the program.
% However, because these qubits become mutually entangled during execution, it raises the concern regarding the correctness of the analysis when they are reused. 
% In fact, we 
% In Section~\ref{}, we will discuss if this entanglement is benign or harmful.

\vspace{5pt}
\noindent \textbf{Looping and Recursion.}
% Unlike classical computers, the state of qubits (superposition) will disappear once they are measured. 
% To determine the number of loops (while or for statements), we have to use fixed point quantum search to amplify the phase of the states which don't satify the constraints and measure the qubits to see if all states in superposition satisfy the loop constraints. 
% However, once we measured the qubits, the superposition will disappear and we have to reproduce the whole circuit, which is extremely time consuming and lose the advantages of quantum computer.
In classical methods, loops and recursive function calls are typically unrolled for a bounded number of iterations, because it is difficult to statically determine the exact iteration count without prior knowledge of all relevant program behaviors.
Quantum techniques might seem to offer a plausible alternative, since qubits representing program states can be mid-circuit measured to determine if the termination condition of a loop or recursion is satisfied and whether another iteration is needed.
However, this approach is over expensive in practice because measuring the qubits causes them to collapse, thereby erasing the program states accumulated thus far.
Thus, the entire circuit must be re-run to reconstruct the state before proceeding, which is extremely time-consuming.

As such, in \sys, we adopt the same heuristics used in classical methods, unrolling all loops a predefined number of times.
For each iteration, \sys duplicates the corresponding circuit component once to encode additional program states onto the same set of qubits.
Although this increases the circuit depth linearly with the number of unrollings, the number of used qubits remains unchanged, as it depends solely on the number of variables and their width. 
Hence, the resource consumption of \sys offers a significant advantage over classical methods such as symbolic execution, where the memory representing program states is forked for each unrolling—one for terminating the loop or recursion and another for continuing the iteration—leading to exponentially growing memory usage.

\subsection{Decoding Program States for Analysis}
In the example program shown in Fig~\ref{fig:ExampleCircuit}, 
when we measure z at the end, the measurement probability of different values of z would be: \begin{equation}
  \begin{aligned}
    P(1)&=\frac{5}{64}, 
    P(2)=\frac{5}{64},
    P(3)=\frac{5}{64},
    P(4)=\frac{5}{64},\\
    P(5)&=\frac{5}{64},
    P(6)=\frac{13}{64},
    P(7)=\frac{13}{64}, 
    P(8)=\frac{13}{64}
  \end{aligned}
\end{equation}

In this expression, the state $|1\rangle$ corresponds to \Code{z} taking the value 1, with a measurement probability of $\frac{5}{64}$.
Put another way, \Code{z} takes the value $1$ in $5$ of all $64$ program states.
By executing the quantum circuit multiple times and measuring the outcomes, we can read out the program states explored by \sys. 
However, the number of measurements required can be enormous, making this approach overly expensive, particularly when dealing with 64-bit wide variables.

Fortunately, in program analysis, analysts are typically interested only in the occurrence of specific states rather than the entire state space.
For example, they may focus on whether \Code{z} can take the value $8$, which indicates an integer overflow in the example program.
Therefore, instead of repeatedly executing the circuit and measuring all possible states, \sys employs the fixed-point quantum search~\cite{fixed-point} to amplify the probability of \Code{z == 8}, allowing for efficient detection of potential overflows.
We do not use Grover algorithm directly, as it requires prior knowledge of the distribution of program states, whereas the fixed-point algorithm does not (See Section~\ref{sec:background}).

% \yc{fixed point is used to examine specific situation of interest. estimate how large N/M is? }
% In \sys, we employ the fixed-point algorithm to amplify the phase of explored program states encoded in qubits, thereby decoding them for analysis.

The time complexity of the fixed-point algorithm is $O(\sqrt{N/M})$, where $N$ is the total number of possible program states, and $M$ is the number of states of interested that are amplified. 
In our example above, $N=64$ represents all the values of two 3-bit variables, while $M=1$ corresponds to the single case, \ie, \Code{x == 1, y == 1}) being amplified.
This offers a quadratic improvement over classical methods designed for the same purpose.
For example, the SAT/SMT solver used in symbolic execution has a time complexity of $O(N-M)$.
However, as our evaluation of \sys will show, the $O(\sqrt{N/M})$ complexity can still be significant in the worst case when $N$ is unbounded and $M$ is $1$.
In Section~\ref{sec:sysh}, we will explore how a hybrid design can integrate classical methods to effectively bound $N$, benefiting program analysis sooner in FTQC.

\begin{figure}[t]
    \centering
	\includegraphics[width=0.5\textwidth]{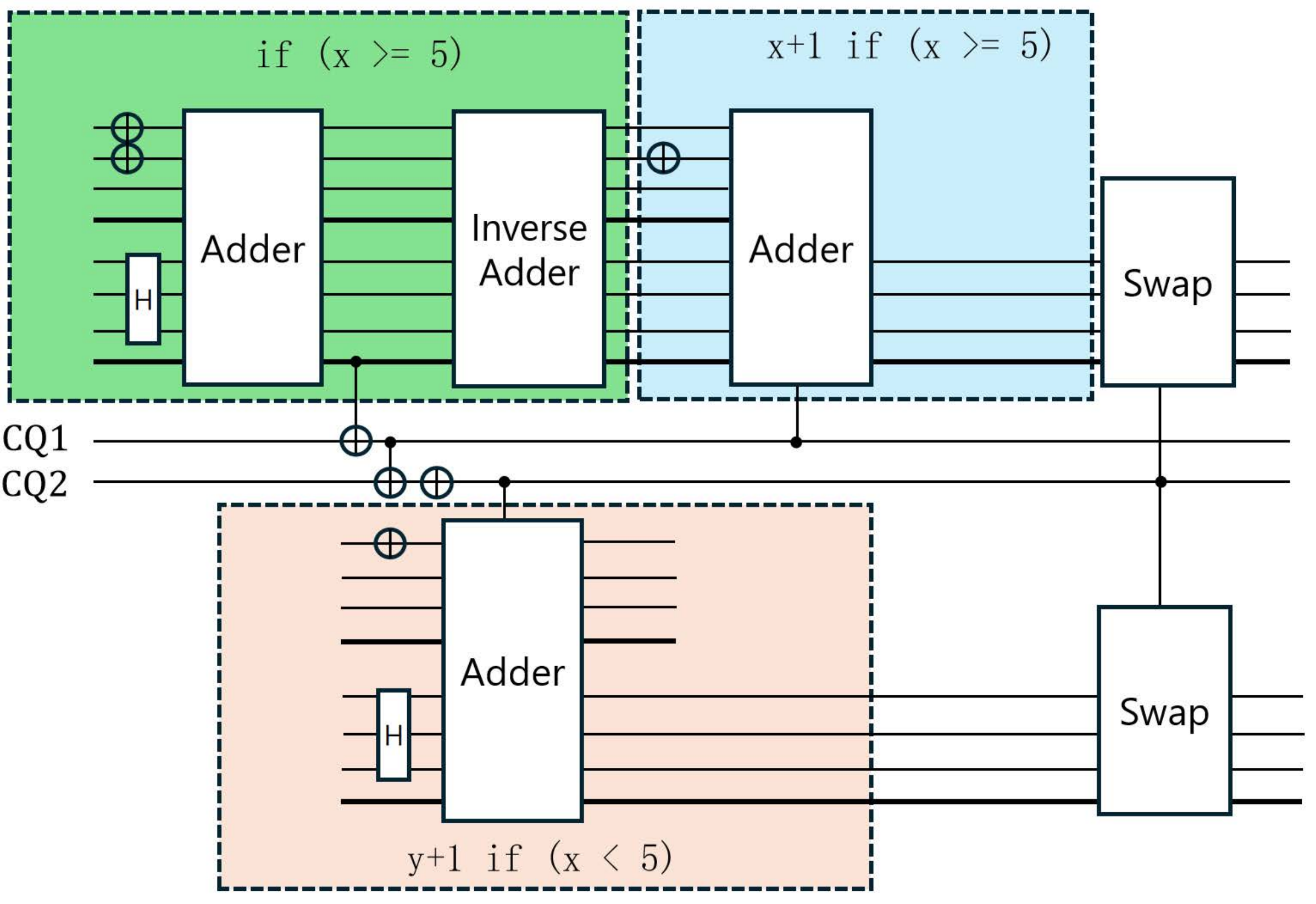}
    \caption{The optimized circuit to interpret the same example program as Figure~\ref{fig:ExampleCircuit}.}
    \label{fig:ExampleCircuit2} 
\end{figure}

\subsection{Optimizations and Trade-offs}
The circuit in Figure~\ref{fig:ExampleCircuit} uses a total of $34$ qubits.
This consumption can be reduced to $16$ qubits by using an inverse adder and sharing qubits among immediate values, as illustrated in Figure~\ref{fig:ExampleCircuit2}.
Beyond this, we can leverage additional qubits for parallelism.
In the following, we discuss these optimizations and trade-offs in detail.

% \todo{This consumption can be reduced to 16 qubits by using an inverse adder and sharing qubits among immediate values , as illustrated in Figure~\ref{fig:ExampleCircuit2}. Beyond this, we can also allocate qubits for parallelism. In the following paragraphs, we will discuss these four optimizations and trade-offs. }

% The circuit in Figure~\ref{fig:ExampleCircuit} uses a total of $34$ qubits.
% This consumption can be reduced to 16 qubits by using an inverse adder, as illustrated in Figure~\ref{fig:ExampleCircuit2} (see Section~\ref{sec:inverse_adder} for details). 
% Additionally, in Sections~\ref{sec_qubit_sharing} and \ref{sec:parallelism} we discuss two optimization schemes that enable trade-offs between qubit usage and circuit depth, offering greater flexibility in resource allocation.
% % #RA: elaborating on the trade-off.  
% % #RA, is this correct? 
% Although this optimization increases circuit depth, in a fault-tolerant regime with quantum error correction (QEC), the impact on fidelity is minimal; the main cost lies in the longer execution time. Consequently, the trade-off between saving qubits and adding depth remains acceptable in most practical settings. 

\begin{figure}[t]
    \centering
	\includegraphics[width=0.8\columnwidth]{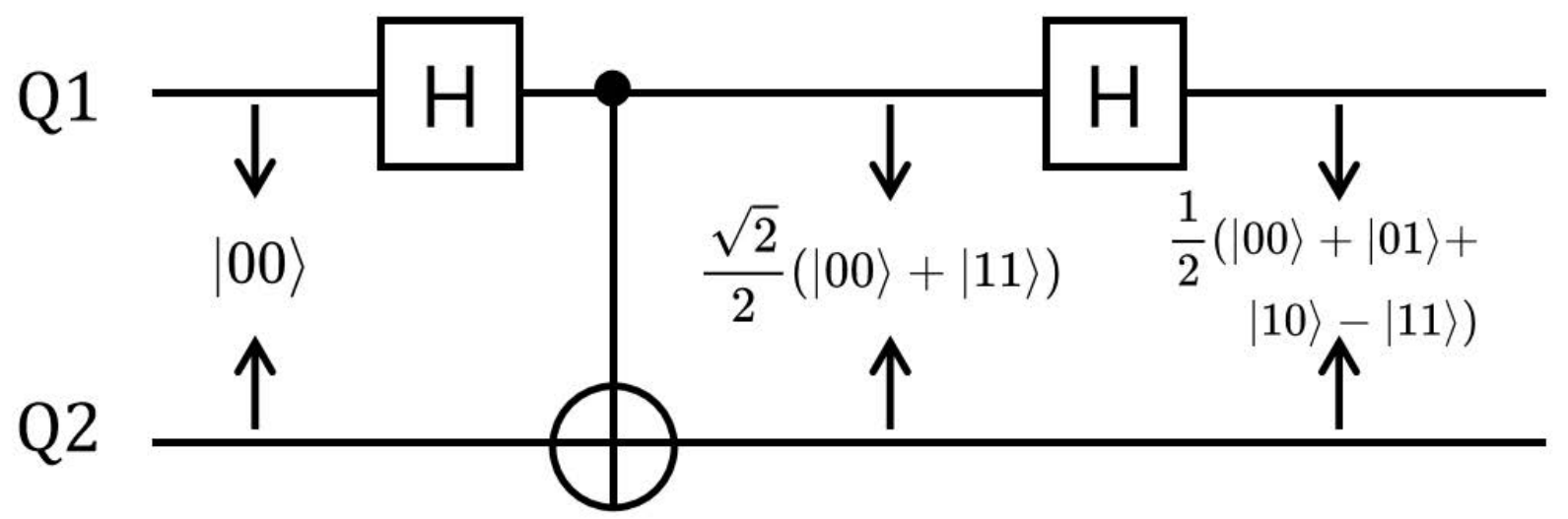}
    \caption{Illustration of problems in phase cancellation.}
    \label{fig:PhaseCancellation} 
\end{figure}

% \noindent \textbf{Original Adder or Extended Adder}
% \todo{As illustrated in Section~\ref{sec:arithmetic}, original adder in paper~\cite{QuantumArithmeticFourier,QuantumArithmeticSignBit} takes two set of qubits as input and one set of qubits represent \Code{x} will be overwritten as \Code{x+1} after the adder. We extend this adder to take three set of qubits as input and store \Code{x+1} on the third set of qubits. However, if \Code{x} will never be used in the future or \Code{x} is a variable needed to be analyzed, we can use original adder to save one set of qubits.}

\vspace{5pt}
\noindent \textbf{Inverse Adder.}
%\subsubsection{Inverse Adder} \label{sec:inverse_adder}
In the unoptimized circuit, a set of four qubits is allocated to interpret \Code{x+3}.
These four qubits, however, are no longer used in subsequent circuit components.
To avoid this waste, we introduce a pair of adder and inverse adder.
The adder sets the signal qubit for predicate evaluation, while the latter inverses \Code{x+3} back to \Code{x}.
This design optimization saves one set of input qubits to the adder while the introduction of inverse adder will increase circuit depth and number of gates. Although this optimization increases circuit depth, in a fault-tolerant regime with quantum error correction (QEC), the impact on fidelity is minimal; the main cost lies in the longer execution time. Consequently, the trade-off between saving qubits and adding depth remains acceptable in most practical settings.
% #RA: here we must immediatly say that increased circuit depth is not a big convern
% copying my comment from below:
% Although this optimization increases circuit depth, in a fault-tolerant regime with quantum error correction (QEC), the impact on fidelity is minimal; the main cost lies in the longer execution time. Consequently, the trade-off between saving qubits and adding depth remains acceptable in most practical settings. 

Another potential problem of such optimization is that complete value reversal may not always be feasible due to entanglement.
For example, in Figure~\ref{fig:PhaseCancellation}, Q1 undergoes an H gate and subsequently entangles with Q2 via a CX gate. 
Attempting to revert Q1 to $|0\rangle$ using another H gate fails because the states $\frac{1}{2}|10\rangle$ and $-\frac{1}{2}|11\rangle$ cannot cancel out due to the entanglement with Q2. 
Fortunately, a sufficient condition for successful phase cancellation is that all values in both the state vector and the inverse matrix are nonnegative.
Since all matrix of arithmetic operations are Boolean and the state vectors throughout the circuit contain no negative value, inversion can be achieved using this optimization.

% As we can see, the set of qubits which represent x+3 will never be used in the continuing circuits. However, we can use the adder U1 instead of adder U2 (Fig \ref{fig:ExampleCircuit2}. 
% Note that x has been updated to x+3 after the first adder. 
% So we need to use inverse adder to inverse x+3 back to x after we copy the sign bit to a control qubit. 
% This could save us one set of qubits , 4 control X gates and 1 circuit depth. 

% However, the problem of phase cancellation may be raised up. In fig \ref{fig:PhaseCancellation}), if we don't create entanglements (without CX gate), qubit 0 will be reversed back to 0. 
% However, due to some entanglements, $\frac{1}{2}|01\rangle$ and $-\frac{1}{2}|11\rangle$ can't be canceled out. 
% One sufficient condition that can guarantee the phase can be successfully canceled out is that the values of state vector after CX gate should be all larger or equal to 0, and the values of inverse matrix after CX gate should also be all larger or equal to 0. 
% And it's easily proved that all the arithmetic operations matrix are constructed by only 1 and 0. 
% So the inverse arithmetic must satisfy the condition. And in our circuit design, the state vectors for all possible stages of the circuit have no negative value. So there's no phase cancellation problem for our algorithm.

\vspace{5pt}
\noindent \textbf{Sharing Qubits among Immediate Values.}
%\subsubsection{Sharing Qubits among Immediate Values} \label{sec_qubit_sharing}
In Figure~\ref{fig:ExampleCircuit}, two immediate values, $3$ and $1$, each consumes a separate set of four qubits.
This can be optimized to reduce the two sets to a single set by applying X gates to transform the value from $3$ to $1$, as shown in Figure~\ref{fig:ExampleCircuit2}.
The same optimization can be applied to the other $1$ in the false branch, allowing all immediate values in the program to share a single set of qubits.
However, this approach increases the circuit depth, as the exploration of the false branch can no longer occur in parallel with the true branch.  
Like using inverse adder, this optimization increases execution time which is often outweighed by the significant qubit savings, particularly within fault-tolerant quantum error correction regimes where fidelity is largely preserved.
% #RA, again, we must highlight that since we will use QEC in FTQC, it is ok to increase depth,, paraphreasing what we said above

% we can see the immediate number can be reused with applying several X gates. 
% So as shown in the example, one set of qubits can represent 3 first and then be converted to 1. 
% We can also use the same set of qubits for the 1 in y+1. 
% However, if we do so, it will increase the circuit depth cause we need to wait for x+1 then do y+1. 
% Parallel speed up will be disabled for qubit reuse. 
% So whether or not to reuse the qubits which represent immediate number not only depends on the specific program we are analyzing, but also the coherence time, gate operation time and qubit number of the current quantum computer.

\vspace{5pt}
\noindent \textbf{Allocating Qubits for Parallelism.}
%\subsubsection{Allocating Qubits for Parallelism} \label{sec:parallelism}
Given the program \Code{if (x >= 5) {z := x + 1;} else {z := x + 2;}}, the two branches must be explored sequentially because the qubits representing \Code{x} are manipulated in both branches. 
However, by rewriting the program as \Code{int y := x; if (x >= 5) { z := x + 1;} else {z := y + 2;}}, a new set of qubits can be allocated to \Code{y}, allowing the branches to be explored in parallel.
This optimization is expected to be used only when qubits are sufficient and shortening execution time is critical.
% trades off increased qubit usage for circuit depth.
% \todo{How do we reframe this part? Ramin may misunderstood this part. In this section we trade qubit usage for circuit depth, not saving qubits.}
% #YC, this part is trading qubits for less circuit depth, so we can't say circuit depth doesn't matter.
% #RA, again, what we gain (saving qubits) is better than increasing cir depth, which we can afford becuase of quantum error correction
% The price of parallel is that we introduce a new set of qubits. 
% This situation is common in the real world programs. Users can generate a balanced circuit to make sure the circuit requires resources within the availability of quantum machine.

\subsection{Is Entanglement Useful or Harmful?}
\label{subsec:engtanglement}
The values that different variables can take in a program are mutually interdependent.
For example, consider the statement \Code{z := x + y}.
If \Code{x} can only be $1$ or $2$ and \Code{y} can only be $3$ or $4$, then \Code{z} can only be $4$, $5$, and $6$. 
In this case, \Code{x == 1} and \Code{z == 6} cannot simultaneously hold.
Such data dependency is beyond the capacity of classical methods like abstract interpretation.
In \sys, the superposition of \Code{x}, \Code{y} and \Code{z} is represented as
\begin{equation}
    \frac{\sqrt{2}}{2}(|1\rangle + |2\rangle)\otimes\frac{\sqrt{2}}{2}(|3\rangle + |4\rangle)\otimes|0\rangle
\end{equation}
When the addition operation is applied, it yields the state 
\begin{equation}
    \frac{1}{2}(|1\rangle|3\rangle|4\rangle + |1\rangle|4\rangle|5\rangle + |2\rangle|3\rangle|5\rangle + \frac{1}{2}|2\rangle|4\rangle|6\rangle)
\end{equation}
which ensures that when we measure \Code{x} to be $|1\rangle$, the measurement of \Code{z} cannot be $|6\rangle$.
Therefore, entanglement inherently preserves data dependency, which is desired in program analysis for achieving accuracy.
% Variables in classical programs have some inner relationship. For example, in this simple statement \texttt{z = x + y}, if possible input range of x is \{1 ,2\}, and possible input range of y is \{3, 4\}. In this case, z would be \{4, 5, 6\}. 
% However, x == 1 and z == 6 can not appear at the same time. 
% For tools like abstract interpretation, it can't represent such relationship between variables. 
% However, our quantum execution can represent this relationship with entanglements. 
% In this example, the superposition of x, y and z is $\frac{\sqrt{2}}{2}(|001\rangle + |010\rangle) \otimes \frac{\sqrt{2}}{2}(|011\rangle + |100\rangle) \otimes |000\rangle$. 
% After the addition, the result will be $\frac{1}{2}|001011100\rangle + \frac{1}{2}|001100101\rangle + \frac{1}{2}|010011101\rangle + \frac{1}{2}|010100110\rangle$. 
% As we can see here, when x is 001, z can't be 110. Entanglements help us create this inner relationship between variables.

\section{Methodology}  \label{sec:method}
In this section, we discuss our evaluation methodology, experimental environment, and benchmarking.
% {\textcolor{red}{TODO: Need the results for hybrid algorithms, including equivalence of QEX and our simulator, impact of range of input, 3 dimensions of hybrid approach and correctness of fixed point. Introduce phase cancellation problems.}}

% \begin{itemize}
%     \item Is the resource consumption within acceptable scale?
%     \item Does \sys{} have a better accuracy or perfermance compared with classical methods?
%     \item Can \hyb{} successfully solve or partially solve obstacles mentioned in section \ref{sec:design}?
% \end{itemize}
% We will first describe our experiment setup and then present our experimental results.

% \subsection{Experiment Setup}
\vspace{0.05in}
\noindent \textbf{Benchmarks and Test Sets.}
For our evaluations, we constructed a test set of $22$ programs and synthesized quantum circuits~\cite{QEXgithub} to analyze them (See Table~\ref{tab:consumption}).
These test cases are selected from two sources. 
One is the benchmark suite of the International Competition on Software Verification (SV-COMP)~\cite{svcomp} which is the largest competition for automated software verification and witness validation. 
Another is programs that implement fundamental algorithms such as GCD and Fibonacci crawled from LeetCode and StackOverflow.
These programs were originally written in C and were slightly modified to be expressed in the WHILE language.

\vspace{0.05in}
\noindent \textbf{Simulation Setup.} 
We employed IBM's Qiskit quantum simulator to execute the synthesized circuits.
However, some synthesized circuits exceeds Qiskit's simulation capabilities.
To address this, we reduce their scale by restricting the width of all program variables to 3 bits.
For programs that still require more than 30 qubits—approximately the maximum simulation capacity of a classical desktop~\cite{qubitsconsumption}—even after this adjustment, we replaced specific quantum circuit components (\eg, quantum adders) with classical equivalents.

More specifically, we enumerated all possible inputs to the program and collected all possible values that the program variables could take.
This simulation is a brute force exploration of the program's state space using classical computer, which is functionally equivalent to \sys.
Table~\ref{tab:quantum_vs_simulator} shows four exemplar test cases comparing the simulation results from both Qiskit and the brute-force exploration. 
The key variable values obtained from the two methods not only form the same set but also follow an identical distribution.
Note that, this brute force exploration is feasible only for the adjusted 3-bit versions of the programs.
Without this adjustment, it would face state explosion issues.

\vspace{5pt}
\noindent \textbf{Figure of Merit.}
We adopt common metrics from classical program analysis to evaluate the accuracy and soundness of \sys.
Given a program variable, a false positive (FP) refers to a value that can never occur in any real execution, whereas a false negative (FN) represents a value that does occur but is not identified by the analysis technique.
Based on these definitions, we calculate the \textbf{over-approximation rate} through
\begin{equation}
\text{over-approximation rate} = \frac{\text{FP \#} + \text{GT \#}}{\text{GT \#}}
\label{eq:over_approximation}
\end{equation}
and the \textbf{under-approximation rate} through
\begin{equation}
\text{under-approximation rate} = \frac{\text{FN \#}}{\text{GT \#}}
\label{eq:under_approximation}
\end{equation}
where GT (ground truth) represents the set of values that can occur during real execution.

\input{tables/equivalence}

% False Positive Rate (FPR) is defined 
% For program variables, c

% We use Fasle Positive/Negative Rate(FPR/FNR) to evaluate the effectiveness of the method.
% {\small 
% \[
%   \mathrm{FPR/FNR} 
%     = \frac{\text{Number of False Positives/Negatives}}{\text{Number of Ground Truth}}
% \]
% }

% \vspace{0.05in}
% For ground truth, we simply try all possible inputs within the input range for each program and collect all possible values of the return variable.

\section{\sys Experimental Results}
\label{sec:evaluation}

In this section, we evaluate \sys in terms of its effectiveness in improving analysis accuracy and soundness as well as its resource consumption.
We first evaluate \sys's effectiveness in eliminating over-approximation and under-approximation compared to classical analysis techniques.
Then, we measure its resource consumption to illustrate its hardware feasibility.

\vspace{5pt}
\noindent \textbf{Over-approximation Rate (Accuracy).}
To evaluate the degree of \sys in eliminating over-approximation and improving analysis accuracy, we compared it with two representative classical analysis techniques: abstract interpretation and symbolic execution. 
Instead of re-implementing these techniques from scratch, we utilized well-established tools—Frama-C for abstract interpretation and Angr for symbolic execution.

For each program in our test set, we focused on the return values of its core functions.
Given the 3-bit version, we are able to enumerate all possible inputs for the function arguments to establish the ground truth. 
We ran the circuit synthesized by \sys for $1000$ times to obtain all states of the return value.
For Angr, we kept feeding the generated constraints to SMT solvers until all states were obtained.
For Frama-C, we considered all values in the integer domain as the analysis results.
The results from all three tools are presented in Table~\ref{tab:fp-rate}.

% This evaluation focuses on all possible values of the return value rather than looking for values satisfy specific constraints, so we don't implement fixed point quantum search here.

\input{tables/fp_rate}

From Table~\ref{tab:fp-rate}, we observe that Frama-C exhibits a high over-approximation rate, exceeding 100\% for most test cases.
In particular, for the \Code{factorial} program, the rate is as high as 170.8\%, indicating that a substantial portion of the program states explored by Frama-C do not occur during real execution.
This excessive over-approximation limits its practical utility in tasks such as program error detection.
Conversely, Angr produced no false positive as the rate is $100\%$.
However, it failed to complete the analysis for $7$ out of $22$ programs due to memory exhaustion. 
Such failure becomes more common when the scale of real-world program increases.
% \todo{These programs are toy problems (most of them are within 100 lines of codes) and in real-world setting more failures can happen.}
% #RA, we may highlight that these are toy problems, and in real-world setting more failes can happen
In contrast, \sys achieves timely completion without false positives, as it encodes only real program states into qubits and leverages quantum superposition to explore multiple states simultaneously, ensuring both efficiency and accuracy.

% To measure the false positive rate of over-approximation, we randomly sampled 15 different inputs and calculated the average. 

% Table~\ref{tab:fp-rate} presents the results of completeness for abstract interpretation, symbolic execution and \sys{}. 
% For the \Code{factorial} program, abstract interpretation exhibited a high false positive rate of 70.8\%, indicating numerous flagged issues that do not manifest during actual execution. 
% Conversely, symbolic execution produced no false positives but failed to complete execution for 7 out of 22 programs due to memory exhaustion. 
% In contrast, the quantum approach achieved timely completion without false positives, as it encodes only real program states into qubits and leverages quantum superposition to explore multiple states simultaneously, ensuring efficiency.

\input{tables/loops}

\vspace{5pt}
\noindent \textbf{Under-approximation Rate (Soundness).} 
\sys shows no false negative in Table~\ref{tab:fp-rate} because all loops and recursions in the test case programs are fully unrolled to a degree that covers every possible program state.
However, determining this ``degree'' requires comprehensive oracles of the program which is impractical and inherently challenging. 
Therefore, when analyzing programs with loops and recursions, \sys can hardly synthesize circuits statically.
To alleviate this limitation, \sys adopts a common solution from classical analysis techniques that unrolls them to a bounded iteration number. 
This solution inevitably leads to miss some program states that only appear beyond the unrolling boundary, resulting in under-approximation similar to classical methods.
Table~\ref{tab:symbolic-quantum} illustrates the loss of soundness when \sys unrolls programs to the same iteration number as Angr for three loop-intensive programs. 
In these cases, \sys exhibits the same under-approximation rate as Angr because both methods fundamentally capture all program states within the unrolling boundary, albeit through different mechanisms—Angr relies on quantifier-free first-order constraints in symbolic execution, whereas \sys leverages quantum superposition.

The cost of unrolling differs significantly between the two approaches. 
In Angr, memory usage grows exponentially since each conditional statement forks program states, doubling memory requirements at each branch. 
In contrast, for \sys, unrolling results in only a linear increase in circuit depth, as each additional unrolling merely duplicates the corresponding circuit components once.

% To assess the degree of under-approximation, we analyzed three loop-intensive programs by unrolling same to the same iteration as symbolic execution 

% Due to the challenge of precisely determining loop iterations, heuristic-based bounded loop unrolling was employed, which may exclude certain program states, leading to false negatives (unsoundness). 
% To assess this impact, we analyzed three loop-intensive programs using the same unrolling bounds as symbolic execution.
% Table~\ref{tab:symbolic-quantum} shows that the quantum approach exhibited an equivalent false negative rate to symbolic execution. 
% This is because symbolic execution, when memory is sufficient, precisely models program states using quantifier-free first-order constraints, whereas the quantum approach achieves comparable accuracy through superposition.

\input{tables/consumption}

\vspace{5pt}
\noindent \textbf{Resource Consumption.}
% \todo{Although \sys is designed for fault-tolerant quantum computer (with enough qubits, long coherence time and high accuracy), we still measure the resources consumption for our testcases to see how far we can realize this algorithm on the real quantum computer.} 
Table~\ref{tab:consumption} presents the resource consumption of the three largest programs in our test set after applying bounded loop unrolling.
When synthesizing circuits for these programs, we favor 
increasing circuit depth to
reducing  qubit usage.
Taking \Code{counting} as an example, it consists of $2,833$ lines of C code, required $788$ logical qubits and $71,171$ quantum gates in the synthesized circuit.
When scaled from the adjusted 3-bit representation back to the original 64-bit representation, the resource demand increased polynomially to $1,088$ qubits and $14,490,048$ gates. 
The growth in resource consumption follows the Table~\ref{tab:consumptionRate} in Appendix~\ref{appendix}.

If the value whose probability to be amplified using the fixed-point algorithm is a 32-bit value with no bounds and only one case is of interest, the worst-case gate consumption can reach the order of $10^{7}\times\sqrt{(2^{32})^{k}}$ where $k$ is the number of input arguments.
For comparison, Shor’s algorithm, which is widely anticipated to break RSA encryption, requires $10,241$ qubits and $2.22\times10^{12}$ gates to factor a 2,048-bit number, with a circuit depth of $1.79\times10^{12}$~\cite{consumption}.
Therefore, \sys and Shor's algorithm are at par in terms of resource consumption, showing its hardware feasibility in FTQC.

%% file: tables/equivalence.tex
\begin{table}[h]
\centering
\caption{Equivalence between Qiskit's simulation results and a brute-force exploration of the program's state space. The number outside the brackets represents key variables' value in the test case, while the number inside the brackets indicates their appearance count, measured from 1,000 Qiskit simulation and across all possible program inputs. Both methods yield similar distributions.}
\renewcommand\arraystretch{1.3}
\resizebox{\columnwidth}{!}{%
\begin{tabular}{l|l|l}
\hline
\textbf{Test Cases} & \textbf{Qiskit} & \textbf{Brute-force Exploration} \\
\hline
closest\_odd 
  & \begin{tabular}[c]{@{}c@{}}%
     1 (248), 3 (244), \\ 5 (256), 7 (252)
    \end{tabular}
  & \begin{tabular}[c]{@{}c@{}}%
     1 (2), 3 (2), \\ 5 (2), 7 (2)
    \end{tabular}\\
\hline
flow\_sensitive
  & \begin{tabular}[c]{@{}c@{}}%
     0 (145), 1 (100), 6 (121),\\ 7 (263), 8 (256), 9 (115)
    \end{tabular}
  & \begin{tabular}[c]{@{}c@{}}%
     0 (1), 1 (1), 6 (1), \\ 7 (2), 8 (2), 9 (1)
    \end{tabular}\\
\hline
fibo\_2calls
  & 0 (500), 1 (500)
  & 0 (1), 1 (1) \\
\hline
afterrec
  & 1 (138), 0 (142), 2 (800)
  & 0 (1), 1 (1), 2 (5) \\
\hline
\end{tabular}
}
\label{tab:quantum_vs_simulator}
\end{table}

%% file: tables/fp_rate.tex
\begin{table}[t]
\centering
\caption{Effectiveness of \sys in eliminating over-approximation, compared with classical techniques—Abstract Interpretation (AI) and Symbolic Execution (SE). The number presented in the table is over-approximation rate.}
\renewcommand\arraystretch{1.2}
\resizebox{\columnwidth}{!}{%
\begin{tabular}{l|r|r|r}
\hline
\textbf{Test Cases}                                             & \multicolumn{1}{c|}{\textbf{AI}} & \multicolumn{1}{c|}{\textbf{SE}} & \multicolumn{1}{c}{\textbf{\sys{}}} \\ \hline
Ackermann01                                        & 115.4\%                      & timeout                  & 100.0\%                         \\ \hline
afterrec                                             & 100.0\%                       & 100.0\%                      & 100.0\%                         \\ \hline
closest\_odd                                           & 145.6\%                      & 100.0\%                      & 100.0\%                         \\ \hline
closest\_prime                                         & 143.8\%                      & 100.0\%                      & 100.0\%                         \\ \hline
counting                                               & 143.8\%                      & 100.0\%                      & 100.0\%                         \\ \hline
divbin                                                & 100.0\%                       & timeout                  & 100.0\%                         \\ \hline
divbin2\_unwindbound5                                  & 143.8\%                      & 100.0\%                      & 100.0\%                         \\ \hline
Et1\_true                                              & 145.8\%                      & 100.0\%                      & 100.0\%                         \\ \hline
factorial                                    & 170.8\%                      & 100.0\%                      & 100.0\%                         \\ \hline
fibo\_2calls                          & 100.0\%                       & 100.0\%                      & 100.0\%                         \\ \hline
fibonacci                                  & 153.8\%                      & 100.0\%                      & 100.0\%                         \\ \hline
flow\_sensitive                                        & 111.4\%                      & 100.0\%                      & 100.0\%                         \\ \hline
gcd                                                   & 117.9\%                      & timeout                  & 100.0\%                         \\ \hline
gcd01                                              & 125.4\%                      & timeout                  & 100.0\%                         \\ \hline
min\_num                                               & 115.1\%                      & timeout                  & 100.0\%                         \\ \hline
nested                                              & 153.1\%                      & 100.0\%                      & 100.0\%                         \\ \hline
nested                                             & 152.4\%                      & timeout                  & 100.0\%                         \\ \hline
num\_conversion                                    & 143.8\%                      & 100.0\%                      & 100.0\%                         \\ \hline
num\_digits\_bin                                       & 166.7\%                      & 100.0\%                      & 100.0\%                         \\ \hline
parity\_transform                                      & 118.2\%                      & 100.0\%                      & 100.0\%                         \\ \hline
prodbin-both-nr                                        & 162.5\%                      & timeout                  & 100.0\%                         \\ \hline
sum\_digits                                            & 143.8\%                      & 100.0\%                      & 100.0\%                         \\ \hline
\end{tabular}
}
\vspace{-2em}
\label{tab:fp-rate}
\end{table}

%% file: tables/loops.tex
\begin{table}[t]
\centering
\caption{Under-approximation rate of \sys and Symbolic Execution (AI) when loops are unrolled to a bounded iteration number.}
\renewcommand\arraystretch{1.2}
\resizebox{0.93\columnwidth}{!}{%
\begin{tabular}{l|c|c|c}
\hline
\textbf{Test Cases}                                 & \textbf{Unroll \#}    & \textbf{SE} & \textbf{\sys{}} \\ \hline
fibonacci           & 2     & 33\%                     & 33\%         \\ \hline
divbin2\_unwindbound & 2     & 0\%                      & 0\%          \\ \hline
counting          & 64     & 71\%                     & 71\%         \\ \hline
\end{tabular}
}
\label{tab:symbolic-quantum}
\end{table}

%% file: tables/consumption.tex
\begin{table}[t]
\centering
\caption{Resource consumption of three largest programs in our test set.}
\renewcommand\arraystretch{1.2}
\resizebox{0.94\columnwidth}{!}{
\begin{tabular}{l|r|r|r|r}
\hline
\textbf{Test Cases} & \textbf{LoC} & \textbf{Qubit \#} & \textbf{Gate \#} & \textbf{Depth} \\ \hline
counting             & 2,833         & 788                                 & 71,171      & 1,060  \\ \hline
fibo\_2calls     & 247          & 204                                   & 4,470         & 1,568                                      \\ \hline
divbin2    & 92           & 68                                   & 16,004     & 11,422                                \\ \hline
\end{tabular}
}
\label{tab:consumption}
\end{table}

%% file: sections/qex-h.tex
\section{\sysh Design and Evaluation}
\label{sec:sysh}

% \sys is designed for FTQC. 
% Our evaluation shows that \sys effectively eliminates both over-approximation and under-approximation compared to classical analysis techniques.
% % #RA, this is not correct
% To make \sys practical in the NISQ era so that program analysis can benefit from it sooner, we introduce a hybrid approach, named \sysh, to enhance the applicability and expedite the adoption of \sys.

% \todo{
Our evaluation shows that \sys effectively reduces over-approximation and under-approximation in program analysis. 
However, \sys cannot well handle language features such as pointer-based operations. 
Additionally, FTQC relies on QEC for high fidelity and reliable operations. 
If we can augment \sys with classical analyses and thus reduce QEC requirements over phyiscal qubits, program analysts are able to benefit from \sys sooner. 
To this end, we introduce a hybrid approach, named \sysh, to enhance the applicability and expedite the adoption of \sys.

\subsection{Motivations of \sysh}

\sys is not able to support syntax features in the WHILE extension. 
In addition to statements described in Section~\ref{sec:design}, the extended WHILE language includes address-of operations (\Code{x := &p}), pointer dereferencing (\Code{x := *p}), and pointer assignments (\Code{*p := x}).
In programs with these pointer-related statements, multiple address-of operators may assign different addresses to the same pointer along distinct program paths, causing the pointer's value to exist in superposition.
Though existing QRAM~\cite{qram0, qram1} supports reading values by dereferencing such superposed pointers, no current design allows writing a superposed value to a superposed address.
Therefore, given the current state of hardware development, \sys is unable to interpret pointer-related statements.

Additionally, hybridizing \sys with classical techniques can reduce QEC requirements in FTQC.  
Our evaluation shows that \sys can incur significant resource consumption in some extreme cases. 
Such scenarios may be rare since in many instances, the parameter $M$ in the time complexity of the fixed-point algorithm, $O(\sqrt{N/M})$, is not $1$ but approaches $N$—considering that an index only needs to exceed a small threshold to overflow a buffer. 
% Additionally, in most cases we don't have an estimation for M and can only assume M is 1 (fixed point quantum search can still keep the accuray even if iterations are more than ideal number). 
However, it remains preferable to generally reduce $N$ by using outcome from classical techniques. 
In this case, we can reduce iteration numbers in the fixed-point algorithm and thus circuit depth and gate usage.
On the other hand, as modern software, for example, the Linux kernel, scales to million lines of code, whole program analysis is not feasible for symbolic execution and abstract interpretation, so is \sys.
Therefore, compositional analysis~\cite{compositional}, which analyzes part of program code, is better suited for \sys, which also necessitates hybridization with classical analysis techniques.

\subsection{\sysh: Hybridizing \sys with Classical Methods}
\sysh doesn't introduce new circuit designs for statement interpretation but leverages the complementary strengths of classical analysis methods to enhance \sys.
In turn, this hybrid approach can also improve classical methods.
In the following, we explore different hybridization strategies and evaluate their effectiveness through case study.

\begin{figure}[t]
    \centering
	\includegraphics[width=0.47\textwidth]{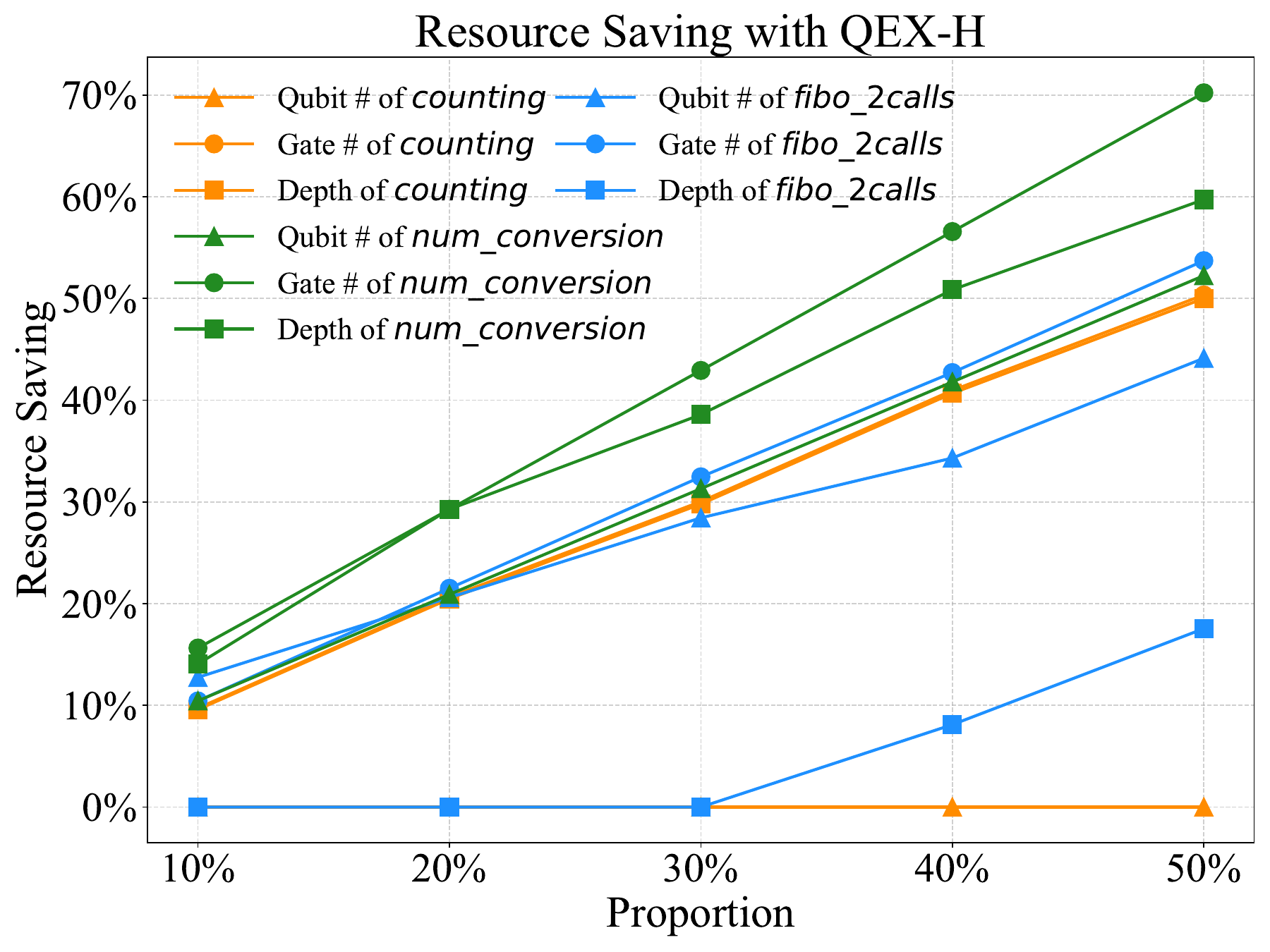}
    \caption{Resource saving with the proportion of codes using classical analysis techniques increased.}
    \label{fig:ResourceSaving} 
\end{figure}

\vspace{5pt}
\noindent \textbf{Straight Hybridization.}
One straightforward hybridization approach is to analyze parts of the program using classical methods, thereby reducing the scale of the quantum circuits and saving resources.
Figure~\ref{fig:ResourceSaving} shows the extent of resource saving as the proportion of classically analyzed parts increases starting from the first line of the program.

From the figure, we observe that the qubit number, gate number, and circuit depth generally decrease linearly as the proportion of classical analysis increases.
An exception is the \Code{counting} case, where all variables are utilized in the latter half of the program, limiting the potential for resource reduction. 
% \todo{The decrease of resource consumption can make our quantum approach realizable sooner in NISQ era.}

\vspace{5pt}
\noindent \textbf{Bound $N$.}
In \sys, the parameter $N$ in the fixed-point algorithm's time complexity can be as high as reach $(2^{32})^{k}$, assuming there are $k$ input arguments, each 32 bit wide.
In \sysh, however, we can bound $N$ by applying \sys to analyze the program's core functions with one or two arguments.
Technically, this is achieved by first using abstract interpretation to analyze the code preceding the core function call.
The resulting over-approximation of the argument’s value range is then provided to \sys.
Though this approximation is not perfectly accurate, it can help bound $N$ and thus reduce the number of iterations and gates. 

Among the $22$ programs in our test set, using the result of Frama-C, $N$ is reduced to the order of $10^{4}$ for $10$ programs: \Code{afterrec-1},  \Code{counting}, \Code{factorial}, \Code{fibonacci}, \Code{gcd}, \Code{nested_1}, \Code{num_conversion_2}, \Code{num_digits_bin}, \Code{parity_transform}, and \Code{sum_digits}.
For another four programs, \Code{Ackermann01}, \Code{Et1_true}, \Code{fibo_2calls}, and \Code{flow_sensitive}, $N$ is bounded below $10$.
In these cases, the gate consumption of \sysh is on the order of $10^{10}$ to $10^{11}$, even fewer than that of Shor's algorithm.
For the remaining eight programs, as part of our future work (See Section~\ref{sec:conclusion}), we plan to optimize the fixed-point algorithm's resource consumption further.

\begin{figure}[t]
\begin{itemize}[left=2em, label={}]
\item
\begin{lstlisting}[caption = {Code snippet includes operations in the extension of WHILE language, illustrating how classical program analysis can mitigate limitations of \sys.} , language = C, label = {lst:example2}]
int func(int x, int* a){ 
    // Analyze with Angr or Frama-C
    *a := x/2;
    int y := x+2;
    int z := *a; 
    // switch to QEX
    if (x>5){ 
        return z*y;
    } else {
        return z+y;
    }
}
\end{lstlisting}
\end{itemize}
\end{figure}

\vspace{5pt}
\noindent \textbf{Skip Pointer-related Statements.}
% Recall that in Section~\ref{}, \sys alone cannot synthesize circuits to interpret address-of operations and pointer assignment in the extension of the WHILE language. 
% This limitation arises from the lack of quantum hardware capable of writing a superpositioned value to a superpositioned address.
For pointer-related statements that cannot be interpreted using quantum circuits due to the limitations of current quantum hardware, \sysh leverages classical techniques to analyze pointer-related operations and then feeds the classical output to \sys. 
This hybrid approach not only overcomes \sys's limitations but also improves analysis accuracy.
% classical analysis techniques for hybridization to overcome this limitation and in the meanwhile preserving accuracy and soundness.

List~\ref{lst:example2} presents a code snippet where pointer-related operations are introduced at lines $3$ and $5$. 
Assuming the input argument \Code{x} can take any value in the integer domain of [0,7], we use Frama-C to perform abstract interpretation on line $2$ to $5$.
After line $5$, the interval domains of variables \Code{y} and \Code{z} are [2,9] and [0,3], respectively. 
If Frama-C continues analyzing the subsequent lines, the final return value would have a domain of [0,27] which contains many false positives.

In contrast, by using the output of Frama-C, we start \sys from line $7$, thereby avoiding $10$ false positives of the return value: $1$, $11$, $13$, $17$, $19$, $20$, $22$, $23$, $25$, $26$.
Despite this improvement, false positives still contain because Frama-C does not preserve data dependencies between variables. 
As a result, the input to \sys is already inaccurate.
For example, \Code{z} can be $3$ on line $5$ only when \Code{y} is $8$ or $9$.
However, \sys mistakenly deems all value pairs of \Code{y} and \Code{z} in the [2,9] and [0,3] domains as valid. 
Alternatively, using Angr to symbolically execute lines $2$ through $5$ preserves data dependencies through first-order logic. 
By consulting SMT solvers, we can obtain all true positive value pairs to feed into \sys, significantly improving accuracy.
% However, as programs scale, individual consulting cannot guarantee the acquisition of all true positives, potentially introducing false negatives to \sys.
% In Section~\ref{}, we will discuss our future plan to directly quantize first-order constraints.

% Intergrating \sys{} with classical methods can also help us save a lot of resources as we only need to run part of the programs. 
% We compute the qubits, gates and depth saving compared with the circuits when fully executed by \sys{}. 
% As we can see in the table \ref{tab:saving}, we randomly select three programs and measure their resource saving rate. 
% We can see all programs have a reduced a large circuit gates and depth when we choose to execute half of the program with \sys{}. 
% For qubits saving, num\_conversion and fibo\_2calls both reach 50\%. Counting has no qubits saving beacause even if it has 50\% codes executed by classical methods, the left 50\% codes will still use all the variables in the program. 
% Different kinds of programs may have different resources saving rate when hybridizing \sys{} with classical methods.

% In sec~\ref{sec:design}, we introduced \hyb{}. Here we make experiments to evaluate the following aspects of \hyb{}:
% \begin{itemize}
    % \item Can \hyb{} improve the constraints for under-constraint symbolic execution?
    % \item Can \hyb{} help deal with memory operations? Are there any side effects?
    % \item Can \hyb{} save resources including qubits, gates and circuit depth?
% \end{itemize}

\begin{figure}[t]
\begin{itemize}[left=2em, label={}]
\item
\begin{lstlisting}[caption = Code snippet of \Code{Ackermann01} to illustrate how \sys can generate accurate inputs to facilitate symbolic execution tool Angr. , language = C, label = {lst:example1}]
int ackermann(int m, int n){
    int res;

    if (m==0) {
        return n+1;
    }
    if (n==0) {
        res = ackermann(m-1,1);
        return res;
    }
    int a = ackermann(m,n-1); // analyze with QEX
    res = ackermann(m-1, a); // analyze with Angr
    return res;
}
\end{lstlisting}
\end{itemize}
\end{figure}

\vspace{5pt}
\noindent \textbf{Generate Accurate Inputs.}
As shown in Table~\ref{tab:fp-rate}, Angr timeouts on $7$ out of $22$ programs due to memory exhaustion.
One classical approach to mitigating this problem is under-constrained symbolic execution~\cite{underconstraint} which directly executes an arbitrary location within the program, effectively skipping the costly path prefix from
\Code{main} function to this location.
However, this approach disregards all constraints imposed by the skipped path (so-called under-constrained), resulting in over-approximation.

In \sysh, we hybridize \sys with Angr by first utilizing \sys to produce accurate constraints for the costly path before commencing Angr.
List~\ref{lst:example1} shows a code snippet from \Code{Ackerman01}, a program on which Angr times out.
Assuming that both arguments \Code{m} and \Code{n} of the \Code{ackermann} function can take on the values \Code{0}, \Code{1}, and \Code{2} at the initial call, line $11$ is executed recursively $16$ times, encountering in total of $42$ \Code{if-else} conditions.
Similarly, line $12$ is executed recursively $16$ times, with $24$ \Code{if-else} conditions in total.
As a result, the total number of branches Angr needs to explore reaches $2^{66}$, which far exceeds the capacity of classical DRAM.

Using under-constrained symbolic execution, we can skip line $11$ and start from line $12$, assuming that \Code{a} can take any value less than $7$ (the maximum representable value in a 3-bit system).
This reduces the number of branches to be explored to $2^{24}$ which is manageable within a 16GB DRAM.
However, this over-approximated input leads to false positives in the final results.
In fact, \Code{m} and \Code{a} only take the following value pairs on line $12$: $(1,2)$, $(1,3)$, $(1,4)$, $(2,3)$, $(2,5)$, $(2,7)$.
To address this, \sysh first quantumly executes line $11$ and feeds the six pairs to Angr to symbolically execute line $12$.
With these accurate inputs, Angr still only needs to explore $2^{24}$ program states while entirely eliminating false positives.

%% file: sections/related.tex
\section{Related Work}
This work explores the potential of leveraging quantum computing for program analysis. 
The most closely related prior works fall into three categories.

\vspace{5pt}
\noindent \textbf{Quantum Solutions to the SAT Problem.}
Solving the Boolean satisfiability problem (SAT) is a fundamental task in various program analysis techniques, including symbolic execution and model checking. 
Several studies have explored quantum approaches to this problem. 
For example, Bian et al.~\cite{SATAnnealer} proposes a method that maps SAT formulas onto D-Wave’s sparse Chimera graph and uses ancillary qubits to handle limited connectivity.
Boulebnane et al.~\cite{SATApproximate} applies the QAOA algorithm to random k-SAT near the satisfiability threshold, providing both a theoretical framework and numerical validation. 
Ayanzadeh et al.~\cite{ayanzadeh2020reinforcement} introduced the Reinforcement Quantum Annealing (RQA) scheme, where an intelligent agent iteratively adjusts Ising Hamiltonians based on feedback from a quantum annealer to enhance the probability of finding global optima in solving SAT.
In the presence of noise and parameter-optimization hurdles, these works discover that quantum approaches may offer advantages for challenging SAT instances. 
By enhancing the capabilities of SAT/SMT solvers, these advancements contribute to the scalability of classical program analysis techniques.
While our work shares a similar objective, it takes a fundamentally different approach.
It leverages quantum superposition to encode program states into qubits and synthesizes quantum circuits to amplify states based on program semantics.
Through the fixed-point algorithm, it readouts the program states of interest according to the analysis goal.

\vspace{5pt}
\noindent \textbf{Analysis of Quantum Programs.} 
As quantum computing is adopted in critical fields, the analysis of quantum programs is becoming increasingly valuable.
Scaffcc~\cite{scaffcc} develops a scalable compiler for large-scale quantum applications to cut overhead and code size.
Kaul et al.~\cite{kaul2023uniform} extends classical code analysis via a Code Property Graph to include quantum-specific information, enabling unified security and correctness checks across both classical and quantum domains. Hung et al.~\cite{hung2019quantitative} extends the existing quantum while-language to handle noisy operations, thereby capturing the reality that quantum gates may err with some probability. It then provides a formal framework to measure and bound the resulting deviation between noisy programs and their ideal, noise-free counterparts, ensuring that developers can quantify and control the effects of quantum noise on program correctness.
Our work takes an opposite direction by applying quantum computing to analyze classical programs.

\vspace{5pt}
\noindent \textbf{High-level Quantum Programming Language.}
To simplify quantum programming, Peter Selinger~\cite{selinger2004towards} introduces a functional quantum programming language that uses classical control to manage quantum data, supporting features like loops, recursion, and structured data types. 
It has a static type system to ensure correctness and a denotational semantics based on complete partial orders of superoperators, providing a high-level abstraction that bridges the gap between quantum circuit models and general-purpose computation.
In addition, Yuan et al.~\cite{QuantumControlMachine} establish which properties of control flow can be correctly implemented on a quantum computer and propose a quantum control machine that uses a restricted conditional jump to realize them. This design provides high-level control flow abstractions via a program counter, avoiding the need to encode all control logic in hardware-level gates. 
In another work, Yuan et al.~\cite{yuan2024t} reduce T-gate costs when abstract control flow for error correction quantum computing because T gates are more expensive than other Clifford gates. 
Those works mainly develop high-level programming languages for quantum computing by incorporating data flow or control flow structures, and thus are different from ours.

%% file: sections/conclude.tex
\section{Conclusion and Future Work}
\label{sec:conclusion}

In this work, we introduce a quantum approach, \emph{\sys}, for program analysis.
\sys leverages superposition to encode program states, enabling the simultaneous exploration of the state space.
Additionally, it utilizes entanglement to track data dependencies, which are crucial for analysis accuracy. 
Our evaluation shows that \sys effectively eliminates over-approximation and under-approximation compared to classical analysis techniques.
To enhance the applicability and scalability of \sys so that it can benefit classical program analysis sooner in FTQC, we propose a hybrid design, \sysh, which integrates \sys with classical techniques.
This hybrid approach supports more language features, reduces resource consumption and maximizes the utility of \sys. 

In the future, we plan to further refine our design to drive its adoption in practice as quantum hardware continues to advance. This effort will involve three key directions.
First, we will develop an algorithm that automatically performs the trade-off between qubit usage and circuit depth according to the characteristics of different program components being analyzed.
Second, we will accelerate the fixed-point algorithm by exponentially searching for the iteration count that ensures convergence of the amplification, leveraging techniques introduced in IQAE~\cite{grinko2021iterative}.
This approach can potentially reduce the iteration number when $M$ is close to $N$, thereby expanding the applicability of our method. Last, we will research on how the probabilistic nature of quantum circuit influence the complexity of our work.

%% file: sections/Appendix.tex
\section{Appendix:Resource consumption for n-bit machine}
\label{appendix}
\subsection{Complete Dirac Representation of Quantum States}
The initial quantum states are given by:
\begin{equation}
|x\rangle = \frac{\sqrt{8}}{8} \sum_{j = 0}^{7} |j\rangle, \quad
|y\rangle = \frac{\sqrt{8}}{8} \sum_{j = 0}^{7} |j\rangle
\label{eq:xy_states} % Optional: Add a label
\end{equation}

The final quantum state for all qubits (except those in pure states) is:

\begin{equation}
\begin{split}
|\phi\rangle =\; &
\frac{1}{8} \sum_{j = 0}^{4} \sum_{k = 0}^{7}
|k+1\rangle\,|j\rangle\,|k\rangle\,|j+3\rangle\,|0\rangle\,|1\rangle \\
& +
\frac{1}{8} \sum_{j = 5}^{7} \sum_{k = 0}^{7}
|j+1\rangle\,|j\rangle\,|k\rangle\,|j+3\rangle\,|1\rangle\,|0\rangle
\end{split}
\label{eq:phi_state}
\end{equation}

Qubits in the first three kets represents \Code{z}, \Code{x} and \Code{y}. Qubits in the fourth ket represents \Code{x+3} (qubits in the bottom of green area). The last two qubits represent CQ1 and CQ2 respectively.

\subsection{Specific Resource Consumption}
Specific resource consumption can't be formalized by $n$ because resource consumption varies significantly from program to program, depending on the specific sequence of operations. However, we can formalize the resource consumption for individual arithmetic operations and \Code{if-else} statements. This analysis is particularly meaningful because, as observed in our three largest test cases (see Table~\ref{tab:consumption}), over 98\% of both gate consumption and circuit depth are attributed to precisely these types of operations. This highlights that arithmetic and conditional logic are often the primary drivers of resource cost in practical quantum programs. Qubit consumption increases linearly with $n$ for arithmetic operations and remains constant for \Code{if-else} statements, so we don't focus on qubit counts here. In contrast, gate consumption and circuit depth exhibit a polynomial increase rate with $n$, as detailed by the formulas in Table~\ref{tab:consumptionRate}. Understanding this polynomial scaling is crucial for predicting the feasibility and performance of algorithms as the number of qubits $n$ increases.
\input{tables/consumptionRate}

%% file: tables/consumptionRate.tex
\begin{table}[b]
\centering
\caption{Gates consumption and circuite depth for five basic operations.}
% 调整表格的行间距和列间距，使单元格更宽敞
\renewcommand{\arraystretch}{1.3} % 行间距默认1.0，适当增大可让每行看起来更宽松
\setlength{\tabcolsep}{10pt}      % 默认约6pt，适当增大可让列间隔更宽

\begin{subtable}[t]{0.4\textwidth}
\centering
\resizebox{.95\linewidth}{!}{%
  \begin{tabular}{l|l}
    \hline
    \textbf{Operation} & \textbf{\# of Quantum Gates} \\ \hline
    Add     & $3n(n+1)/2$ \\ \hline
    Sub     & $3n(n+1)/2$ \\ \hline
    Mul     & $(11n^3-16n^2+5n)/2$ \\ \hline
    Div     & $n(28n^2 + 4n + 4)$ \\ \hline
    If-else & $9n(n+1)+1$ \\ \hline
  \end{tabular}%
}
\end{subtable}

\vspace{0.3cm}

\begin{subtable}[t]{0.4\textwidth}
\centering
\resizebox{.95\linewidth}{!}{%
  \begin{tabular}{l|l}
    \hline
    \textbf{Operation} & \textbf{Circuit Depth} \\ \hline
    Add     & $5n-2$ \\ \hline
    Sub     & $5n-2$ \\ \hline
    Mul     & $(11n^3-18n^2+9n)/2$ \\ \hline
    Div     & $22n^3 + 3n^2 + 6n + 1$ \\ \hline
    If-else & $10n-3$ \\ \hline
  \end{tabular}%
}
\end{subtable}

\label{tab:consumptionRate}
\end{table}

% \begin{table}[t]
% \centering
% \begin{subtable}[t]{0.5\textwidth}
% \centering
% \resizebox{\linewidth}{1.2cm}{%
%   \begin{tabular}{|l|l|}
%     \hline
%     \textbf{Type} & \textbf{\# of Quantum Gates} \\ \hline
%     Add     & $3n(n+1)/2$ \\ \hline
%     Sub     & $3n(n+1)/2$ \\ \hline
%     Mul     & $(11n^3-16n^2+5n)/2$ \\ \hline
%     Div     & $n(58n^2-n+5+2^{2n+1}+2^n)/2 + 2^n - 1$ \\ \hline
%     If-else & $9n(n+1)+1$ \\ \hline
%   \end{tabular}%
% }
% \end{subtable}

% \vspace{0.3cm}

% \begin{subtable}[t]{0.5\textwidth}
% \centering
% \resizebox{\linewidth}{1.2cm}{%
%   \begin{tabular}{|l|l|}
%     \hline
%     \textbf{Type} & \textbf{Depth} \\ \hline
%     Add     & $5n-2$ \\ \hline
%     Sub     & $5n-2$ \\ \hline
%     Mul     & $(11n^3-18n^2+9n)/2$ \\ \hline
%     Div     & $20n^3+4n^2+6n+n\,2^n(2^n-1)/2+1$ \\ \hline
%     If-else & $10n-3$ \\ \hline
%   \end{tabular}%
% }
% \end{subtable}

% \caption{Gates consumption and circuite depth for five basic operations}
% \label{tab:consumptionRate}
% \end{table}